\documentclass[twoside,twocolumn]{article}

\usepackage{blindtext} 

\usepackage[sc]{mathpazo} 
\usepackage[T1]{fontenc} 
\usepackage{microtype} 

\usepackage[english]{babel} 

\usepackage[hmarginratio=1:1,top=32mm,columnsep=20pt]{geometry} 
\usepackage[hang, small,labelfont=bf,up,textfont=it,up]{caption} 
\usepackage{booktabs} 

\usepackage{lettrine} 

\usepackage{enumitem} 
\setlist[itemize]{noitemsep} 

\usepackage{abstract} 

\usepackage{titlesec} 
\renewcommand\thesection{\Roman{section}} 
\renewcommand\thesubsection{\roman{subsection}} 
\titleformat{\section}[block]{\large\scshape\centering}{\thesection.}{1em}{} 
\titleformat{\subsection}[block]{\large}{\thesubsection.}{1em}{} 
\newcommand{\furl}[1]{\footnote{\scriptsize \url{#1}}}
\newcommand{\quotes}[1]{``#1''}

\usepackage{fancyhdr} 
\usepackage{titling} 

\usepackage{hyperref} 
\usepackage{float}
\usepackage{subfig}
\usepackage{url}
\usepackage{comment}
\usepackage{graphicx}
\usepackage{grffile}

\pretitle{\begin{center}\huge\bfseries} 
\title{SDM-RDFizer: An RML Interpreter for the Efficient Creation of RDF Knowledge Graphs}

\author{%
\textsc{Enrique Iglesias}\thanks{The three authors contributed equally to this research.}\\[1ex] 
\normalsize L3S Research Center Leibniz University of Hannover \\ 
\normalsize {s6enigle@uni-bonn.de} 
\and 
\textsc{Samaneh Jozashoori}\footnotemark[1]\\[1ex]
\normalsize TIB Leibniz Information Center for Science and Technology\\ 
\normalsize {samaneh.jozashoori@tib.eu} 
\and 
\textsc{David Chaves-Fraga}\footnotemark[1]\\[1ex] 
\normalsize Ontology Engineering Group - Universidad Politécnica de Madrid\\ 
\normalsize {dchaves@fi.upm.es} 
\and 
\textsc{Diego Collarana}\\[1ex]
\normalsize Fraunhofer IAIS \& University of Bonn\\ 
\normalsize {collaran@cs.uni-bonn.de} 
\and 
\textsc{Maria-Esther Vidal}\\[1ex] 
\normalsize TIB Leibniz Information Center for Science and Technology\\ 
\normalsize {maria.vidal@tib.eu} 
}
\date{} 
\renewcommand{\maketitlehookd}{%
\begin{abstract}
\noindent IIn recent years, the amount of data has increased exponentially, and knowledge graphs have gained attention as data structures to integrate data and knowledge harvested from myriad data sources. However, data complexity issues like large volume, high-duplicate rate, and heterogeneity usually characterize these data sources, being required data management tools able to address the negative impact of these issues on the knowledge graph creation process.  In this paper, we propose the SDM-RDFizer, an interpreter of the RDF Mapping Language (RML), to transform raw data in various formats into an RDF knowledge graph. SDM-RDFizer implements novel algorithms to execute the logical operators between mappings in RML, allowing thus to scale up to complex scenarios where data is not only broad but has a high-duplication rate. We empirically evaluate the SDM-RDFizer performance against diverse testbeds with diverse configurations of data volume, duplicates, and heterogeneity. The observed results indicate that SDM-RDFizer is two orders of magnitude faster than state of the art, thus, meaning that SDM-RDFizer an interoperable and scalable solution for knowledge graph creation. SDM-RDFizer is publicly available as a resource through a \href{https://github.com/SDM-TIB/SDM-RDFizer}{Github repository} and a \href{https://doi.org/10.5281/zenodo.3872103}{DOI}.

\end{abstract}
}

\begin{document}

\maketitle


\section{Introduction}
Knowledge graphs have gained momentum as data structures to integrate--as factual statements-- data and knowledge present in heterogeneous data sources. 
DBpedia and Wikidata are exemplary encyclopedic knowledge graphs frequently accessed by scientific and industrial communities; e.g., only Wikidata receives billions of visits per year\furl{https://stats.wikimedia.org/}. 
Similarly, knowledge graphs are receiving significant attention in science and industrial developments \cite{AuerKPKSV18,NoyGJNPT19}. In fact, according to Google, knowledge graph is a trend term\furl{https://trends.google.com/trends/explore?q=knowledge\%20graph} and the Google Scholar indexes more than 3,5M entries of scientific publications with the term knowledge graph.
These results demonstrate the success in the adoption of Semantic Web technologies. But, they also put into perspective the need to provide efficient and mature techniques for creating and maintaining knowledge graphs. 

Diverse approaches have been proposed to define the process of integrating heterogeneous datasets into knowledge graphs \cite{chebotko2009semantics,calvanese2017ontop,Chaves-FragaEIC19,PriyatnaCS14}. Mapping languages (e.g., R2RML \cite{das2012r2rml} and RML \cite{DimouSCVMW14}) and engines (e.g., RMLMapper\furl{https://github.com/RMLio/rmlmapper-java} and RocketRML \cite{csimcsek2019rocketrml}) represent valuable contributions for performing this transformation process. As observed in our next example, albeit highly used, existing approaches lack efficient data management techniques demanded to create knowledge graphs from large and heterogeneous datasets with duplicates.     

\begin{figure*}[t!]
\centering
\includegraphics[width=0.95\textwidth]{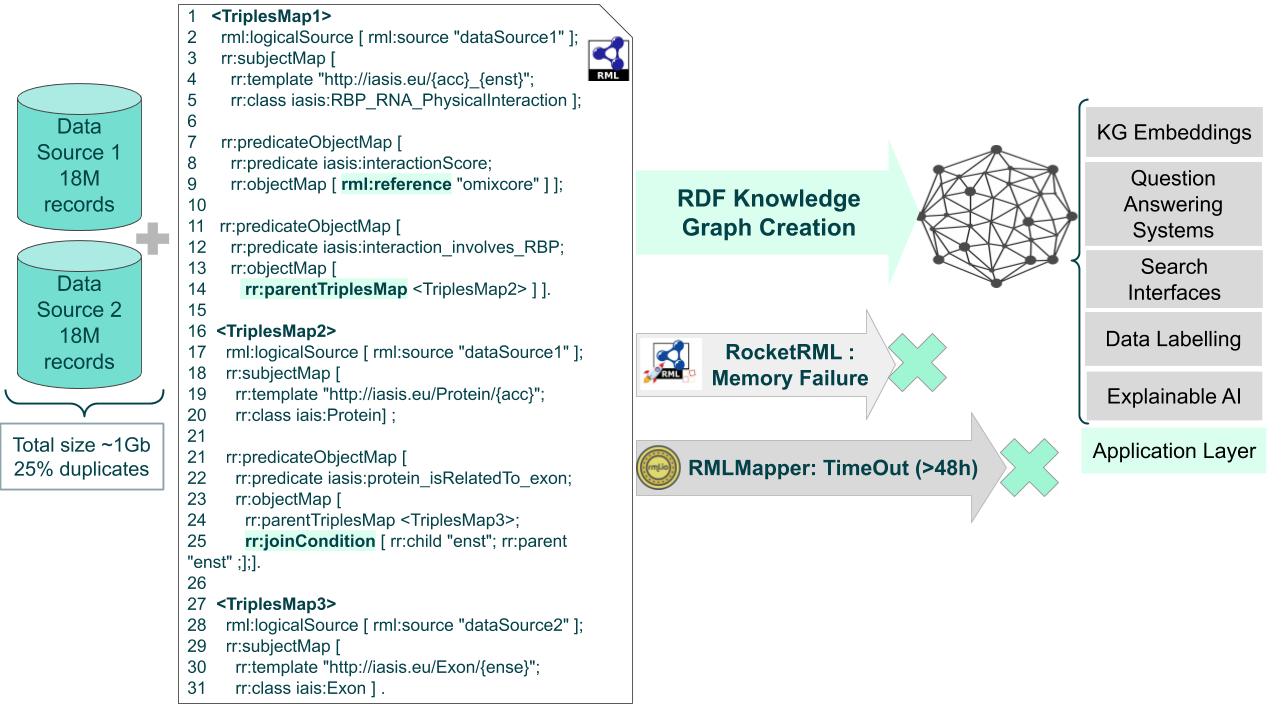}
\caption{\textbf{Motivating example.} Available RML engines, implementing naive join strategy, fail to create a KG from two biomedical datasets with a total size of 1GB and 25\% duplicates.} 
\label{fig:motivatingExample}
\end{figure*}

\noindent \textbf{Motivating Example:}
Creating a knowledge graph from biomedical data sources is an exemplary scenario of being overwhelmed by the volume and heterogeneity of data. 
In~\autoref{fig:motivatingExample}, we see a normal process of transforming two real-world data sources into an RDF knowledge graph using an available RML interpreter. 
In this example, the aim is to integrate data related to the biological concept RBP\_RNA\_PhysicalInteraction\footnote{Protein(RBP)-RNA binding interactions are shown to play essential roles in diseases.
Although there is a lack of enough experimental data, various computational methods are filling this gap by predicting physical interactions between RBP and target RNAs.} from different sources into RDF. 
Accordingly, the subject of \texttt{TripleMap1} represents the mentioned concept. 
Because related data is residing in two different sources, a \textit{Join Condition} is applied in the mapping rules to create the required triples.
It should be noted that even though only four different attributes of both data sources are utilized, the data volumes are considerably large; about 1 Gigabyte in total. 
In this example, to transform the raw data into RDF, two widely accepted RML-compliant interpreters\furl{https://github.com/RMLio/rml-implementation-report}, i.e., RMLMapper\furl{https://github.com/RMLio/rmlmapper-java} and RocketRML~\cite{csimcsek2019rocketrml} are executed. 
However, none of the mentioned engines accomplish the task. 
RocketRML stops early due to the failure of the memory capacity, while RMLMapper times out after 48 hours. 
This example reveals the need for developing mapping engines able to scale up to complex scenarios.
\\
\textbf{Our Resource:} We address the problem of efficient knowledge graph creation and propose a resource named SDM-RDFizer, which can transform data from myriad data sources into an RDF knowledge graph. SDM-RDFizer implements a set of unique physical operators and data structures that speed up the execution of the mapping rules that specify a knowledge graph creation process. The current version of SDM-RDFizer is customized for RML, a mapping language extensively used to create knowledge graphs in diverse domains~\cite{DimouSCVMW14}.    
SDM-RDFizer is publicly available as a resource in a Github\furl{https://github.com/SDM-TIB/SDM-RDFizer} and in Zenodo\furl{https://doi.org/10.5281/zenodo.3872103}. SDM-RDFizer is used in more than eight international projects. Moreover, experimental results reveal the contribution that SDM-RDFizer makes to the repertoire of efficient technologies for knowledge graph management. \\
This paper is structured in six additional sections. Section \ref{sec:pre} summarizes preliminaries, \autoref{sec:approach} defines the data management techniques implemented in SDM-RDFizer, and \autoref{sec:resource} describes SDM-RDFizer as a resource. The results of the empirical evaluations are reported in \autoref{sec:eval}, while the state of the art is summarized in \autoref{sec:rw}, and \autoref{sec:conclusion} wraps up and outlines future work.    
  
\begin{figure*}[t!]
    \centering
    \includegraphics[width=1\textwidth]{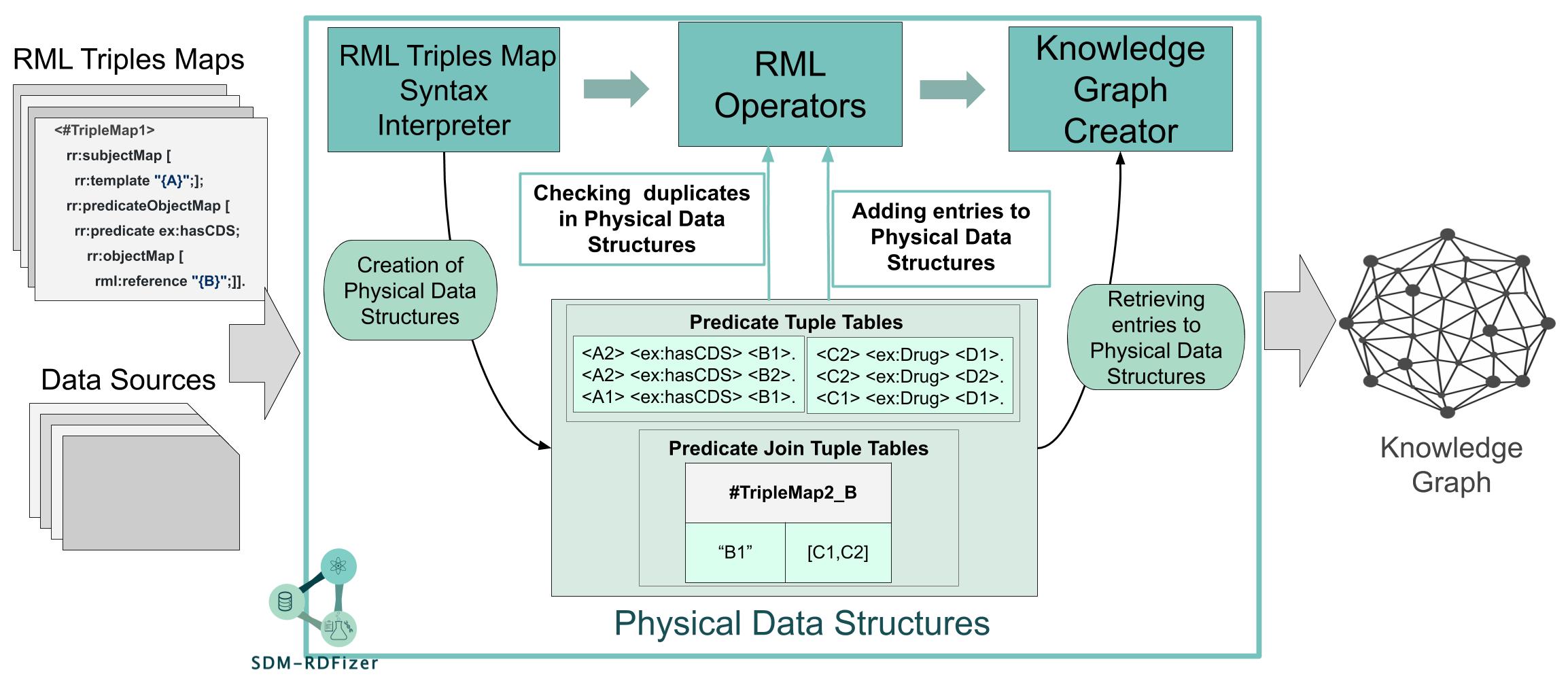}
    \caption{The architecture of the SDM-RDFizer.}
    \label{fig:architecture}
\end{figure*}
\section{Preliminaries}
\label{sec:pre}
\subsection{RDF Mapping Language - RML}
RML~\cite{DimouSCVMW14} is a declarative mapping language used to transform heterogeneous data, including JSON, CSV, and XML, into the RDF data model. 
In RML, we write a customized set of rules expressed under the RML vocabulary\furl{https://rml.io/specs/rml/}.
These rules define mapping rules between entities and properties, from data structures to RDF triples, i.e., RML \emph{Triples Map}s.
Figure~\ref{fig:motivatingExample} illustrates fragments of RML triples maps from our project iASiS\furl{https://www.project-iasis.eu/}. 
First, a \emph{Triples Map} tag groups the mappings sharing the same subject (lines 1, 16, and 27).
Then, we define the data source by defining a \texttt{rml:logicalSource} element (line 2, 17, and 28).
Next, we use the \texttt{rr:subjectMap} element to define the class of the subject.
For example, to map entities in \quotes{dataSource2} to the class \texttt{iasis:Exon}, we use a template-valued term map and the RML class property (lines 29-31).
Afterward, a set of predicate-object maps (\texttt{rr:predicateObjectMap}) elements define the creation of predicates and its object value for the RDF subject.
Lines 7 to 9 show the creation of a \texttt{iasis:interactionScore} predicate referencing the property \quotes{omixcore} from dataSource1. Lines 11 to 14 show the creation of triples when its object is the subject of another triples map, and the same source is used.
Lastly, to integrate data from different sources, it is used as a reference through a join (\texttt{rr:joinCondition}) to another triples map (lines 21-25).
Once we have defined the mapping rules, an RML interpreter performs the data translation across the sources to RDF.
Different characteristics of the data impact the performance of an RML interpreter.
For example, to integrate different sources, several object-join triples maps between data sources are required.
In the following sub-section, we describe the main parameters affecting an RML interpreter.
\subsection{Parameters affecting an RML Interpreter}
Chaves-Fraga et al.~\cite{Chaves-FragaEIC19} empirically analyze the parameters affecting the performance of an RML interpreter.
Five dimensions are defined considering different complexity variables, e.g., mapping, data, platform, source, and output dimensions.
The \emph{data dimension} takes different data characteristics into account, including the dataset size, the data frequency distribution, the partitioning type, and the data format.
The \emph{platform dimension} considers the hardware resources used during knowledge graph creation.
The \emph{source dimension} defines variables related to the access of the data source such as the data source transfer time, limitation in the access, which affect the creation time.
Finally, the \emph{output dimension} defines three variables, i.e., the serialization type of the output, whether the interpreter removes duplicates or not, and the generation type generating the output at once or in a streaming manner.
From an interpreter perspective, Chaves-Fraga et al.~\cite{Chaves-FragaEIC19} show that the dimensions of mapping complexity, data, and output are the ones that impact the most.
Efficient join operators are of paramount importance for an RML interpreter efficiency during execution time.
Further, removing duplicates efficiently during the transformation process positively affects the performance of an RML interpreter.
\section{The SDM-RDFizer: An RML Engine}
\label{sec:approach}
This section describes the SDM-RDFizer in terms of its architecture, the physical operators that make up the execution engine of RML triples maps, and the main properties of these operators.
\begin{figure*}[t!]
 \centering
 \subfloat[Predicate Tuple Table]{
      \includegraphics[width=0.9\columnwidth]{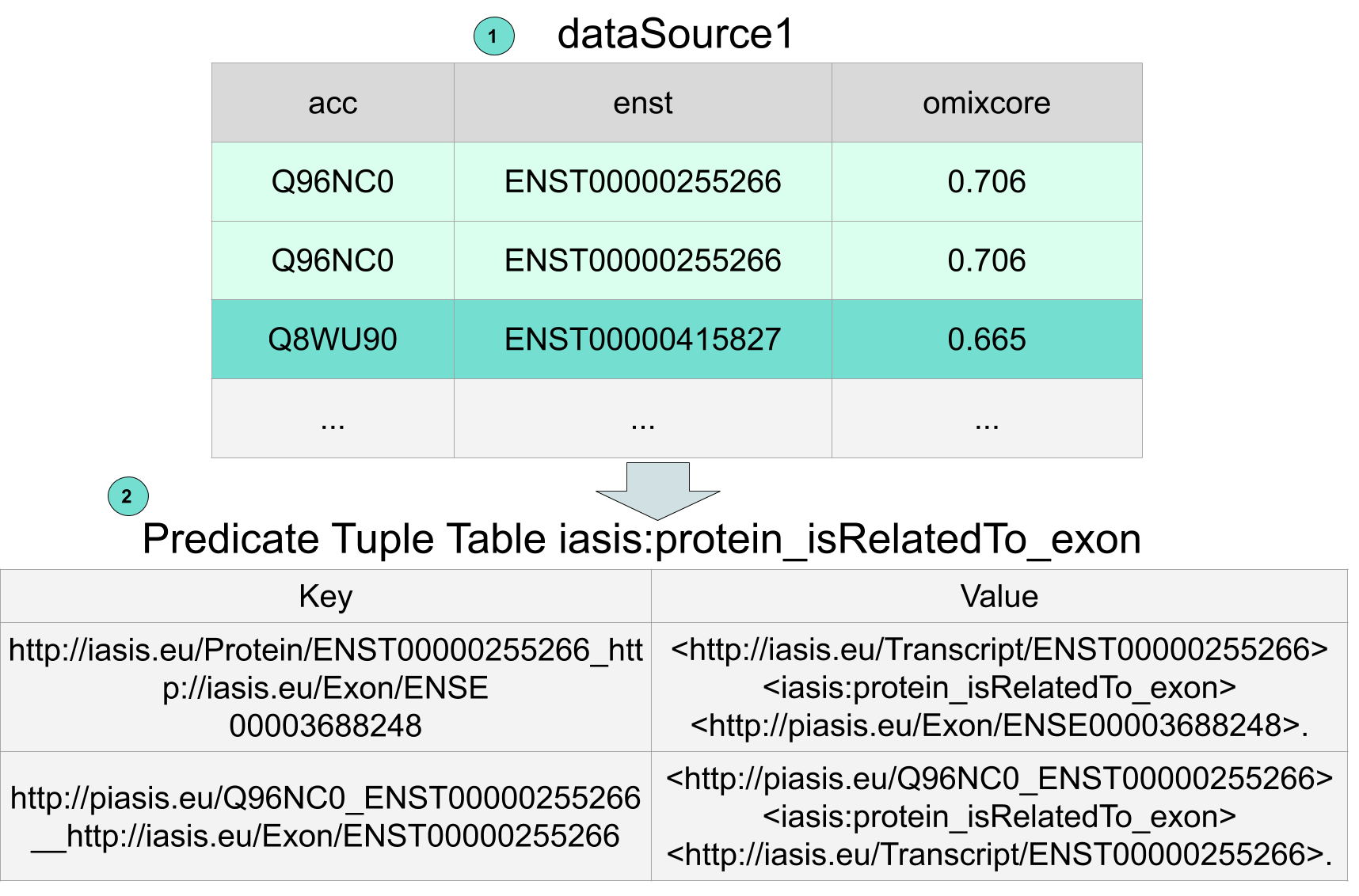}
    \label{fig:ptt}}
  \subfloat[Predicate Join Tuple Table]{
\includegraphics[width=0.85\columnwidth]{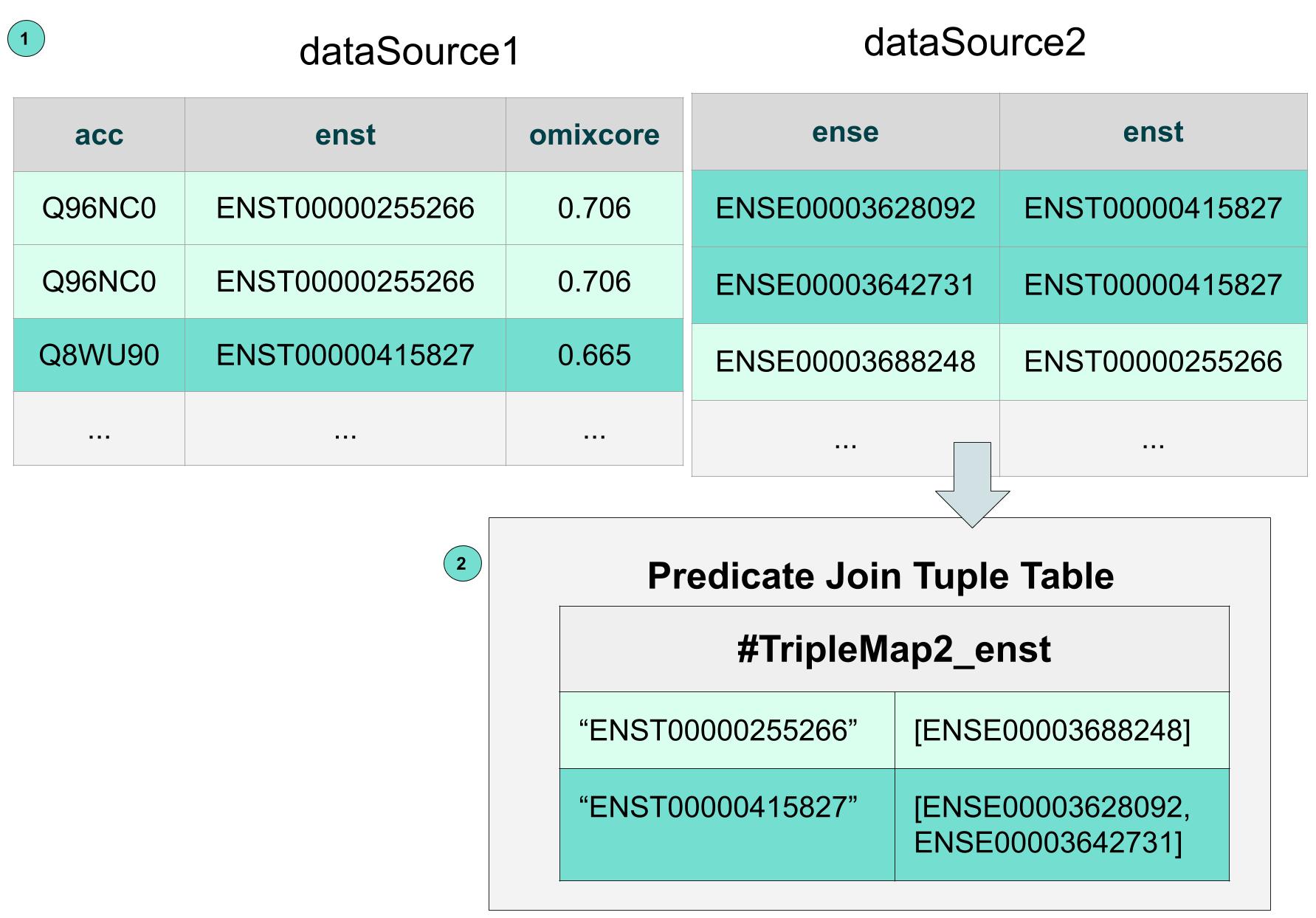}
    \label{fig:pjtt}}
    \caption{{\bf The Physical Data Structures.} The two physical data structures used by SDM-RDFizer are illustrated. (a) A Predicate Tuple Table with three entries. (b) A Predicate Join Tuple Table with two entries.}
    \label{fig:hash_table}
\end{figure*}
\subsection{The SDM-RDFizer Architecture}
A data integration system \textit{DI} provides an abstract representation $\langle O,S,M\rangle$ to specify the mapping rules in a set $M$ that define the integration of a set $S$ of data sources into instances of a unified schema or ontology $O$. 
SDM-RDFizer receives as input a data integration system \textit{DI} and produces as output instances of the concepts in $O$ that result from the execution of the mapping rules in $M$ over the data sources in $S$. 
The current version of SDM-RDFizer is customized for interpreting data integration systems where mapping rules are specified in RML, and the output corresponds to an RDF knowledge graph with the ontology $O$. 
However, the SDM-RDFizer architecture can be easily implemented for other RDF based mapping languages (e.g., R2RML~\cite{das2012r2rml}).  
The execution of the RML triples maps requires the interpretation of triples maps, the creation of physical data structures to store the results of the execution of the RML rules, and the generation of the knowledge graph from the results stored in the data structures. \autoref{fig:architecture} depicts SDM-RDFizer in terms of four main components that implement these steps. 

\noindent\textbf{RML Triples Map Syntax Interpreter:} translates the RML triples maps into the physical data structures that SDM-RDFizer uses to execute the RML triples maps and generate the RDF triples.

\noindent\textbf{RML Operators:} execute the interpreted triples maps over the respective data sources to generate RDF triples. 
During the execution of these operators, the physical data structures are accessed to check if an RDF triple has already been created.
If so, the generation of a duplicated RDF triple is avoided; otherwise, the triple is stored in the physical data structure.
SDM-RDFizer has three operators; they are explained in more detail in \autoref{operators}.
    \begin{itemize}
        \item \textit{Simple Object Map}: is the most basic of the operators to evaluate a \textit{simple predicate object map} statement in an RML triples map. The values of the \textit{object values} are collected from an attribute in the triples map source or are a constant. In the motivating example, this operator generates RDF triples according to the predicate object map in lines 7-9 in \autoref{fig:motivatingExample}.  
        \item \textit{Object Reference Map}: this operator \textit{references a second triples map}. The object of the first triples map is the subject of the second triples map. The main condition for this operator to work is that both triples maps have the same data source. An application of this operator in the motivating example can be seen in lines 11-14. 
        \item \textit{Object Join Map}: this operator executes a \textit{join condition} between two RML triples maps with different data sources. In the motivating example, this operator is utilized to execute the predicate object map in lines 21-25 in \autoref{fig:motivatingExample}.
    \end{itemize}
    
\noindent\textbf{Physical Data Structures:} store results generated so far and avoid the generation of duplicates during the execution of RML triples maps. They are of two types: i) Predicate Tuple Table (PTT): stores per each of predicate $p$ in at least one triple map, the RDF triples generated for $p$ so far. ii) Predicate Join Tuple Table (PJTT): stores the values of the subjects generated by a triples map that is associated with the values that meet a join condition in the triples map. These structures are explained in more detail in \autoref{pds}.

\noindent\textbf{Knowledge Graph Creator:} collects RDF triples stored in PTTs and adds them to the output knowledge graph. The knowledge graph creation is performed incrementally, i.e., as soon as a new RDF triple is added into a PTT, the RDF triple is also included in the knowledge graph. To avoid the same RDF triple to be added more than once, the knowledge graph creator maintains per PTT $t$, the timestamp of the last RDF triple that was selected from $t$. 

\subsection{Physical Data Structures}
The SDM-RDFizer utilizes two physical data structures as a means to optimize the creation of knowledge graphs. These data structures help remove duplicates and avoid unnecessary operations, like uploading the parent triples map's data source of a join multiple times. In the following subsection, the physical data structures used by SDM-RDFizer are described. 
\label{pds}

\noindent\textbf{Predicate Tuple Table (PTT)} 
For each predicate $p$ defined in an object triples map, a PTT is created to store the RDF triples generated so far. Physically, PTTs are implemented as hash tables where the hash key of an entry corresponds to an encoding of the subject and object of a generated RDF triple, and the value of the entry corresponds to the RDF triple. The primary use of this table is to avoid the duplicate generation of an RDF triple. 
If a generated RDF triple is present within PTT, that means that the triple has been previously created and needs to be discarded. But if the generated RDF triple is not within PTT, then it is new and must be added to PTT and the knowledge graph. As shown in Figure\autoref{fig:ptt}, the data source is transformed into RDF triples, which are checked in the corresponding PTT. As RDF triples of a predicate, $p$ can be generated from the execution of different triples maps. PTTs bring significant savings not only in sources with high-duplicated rates but also when data sources create RDF triples of $p$ also overlap.  

\noindent
\noindent\textbf{Predicate Join Tuple Table (PJTT)}
A PJTT stores the values generated during the execution of a join condition between two RML triple maps, e.g., lines 21-25 in \autoref{fig:motivatingExample}, the predicate object map is defined in terms of a join of triples map \texttt{TriplesMap1} (child map) to \texttt{TriplesMap2} (parent map). For each RML triples map $M_i$ that is referred as a parent triples map in a join condition $B$, a predicate join tuple table $M_i \_B$ is created, e.g., \texttt{TriplesMap2\_enst} in our running example. 
Physically, predicate join tuples are hash tables. The hash key of an entry corresponds to the encoding of each of the values of the attributes in the condition $B$ (e.g., \texttt{enst}). Further, the value of the entry is a set with all the subject values generated by $M_i$ (e.g., values of the subject of \texttt{TriplesMap2}) that are associated with the values of the attributes in $B$ represented in the entry hash key.  
 Additionally, a PJTT enables direct access to the subjects associated with the join condition $B$, allowing thus for the join implementation as an index join.  
 In the example shown in Figure\autoref{fig:pjtt}, "enst" is the join condition between the triples maps. The data is organized as the values of the join conditions with its respective value in dataSource2. For example, we have the value "ENST00000415827" and its associated values "ENSE00003628092" and "ENSE00003642731".  In PJTT, "ENST00000415827" is the key in the hash table and "ENSE00003628092" and "ENSE00003642731" are the values. Finally, to identify an entry in PJTT, a key is generated from the identifier of the parent triples map and the join condition.  
\begin{figure*}[t!]
 \centering
 \subfloat[Simple Object Map]{
      \includegraphics[width=0.65\columnwidth,height=6.9cm]{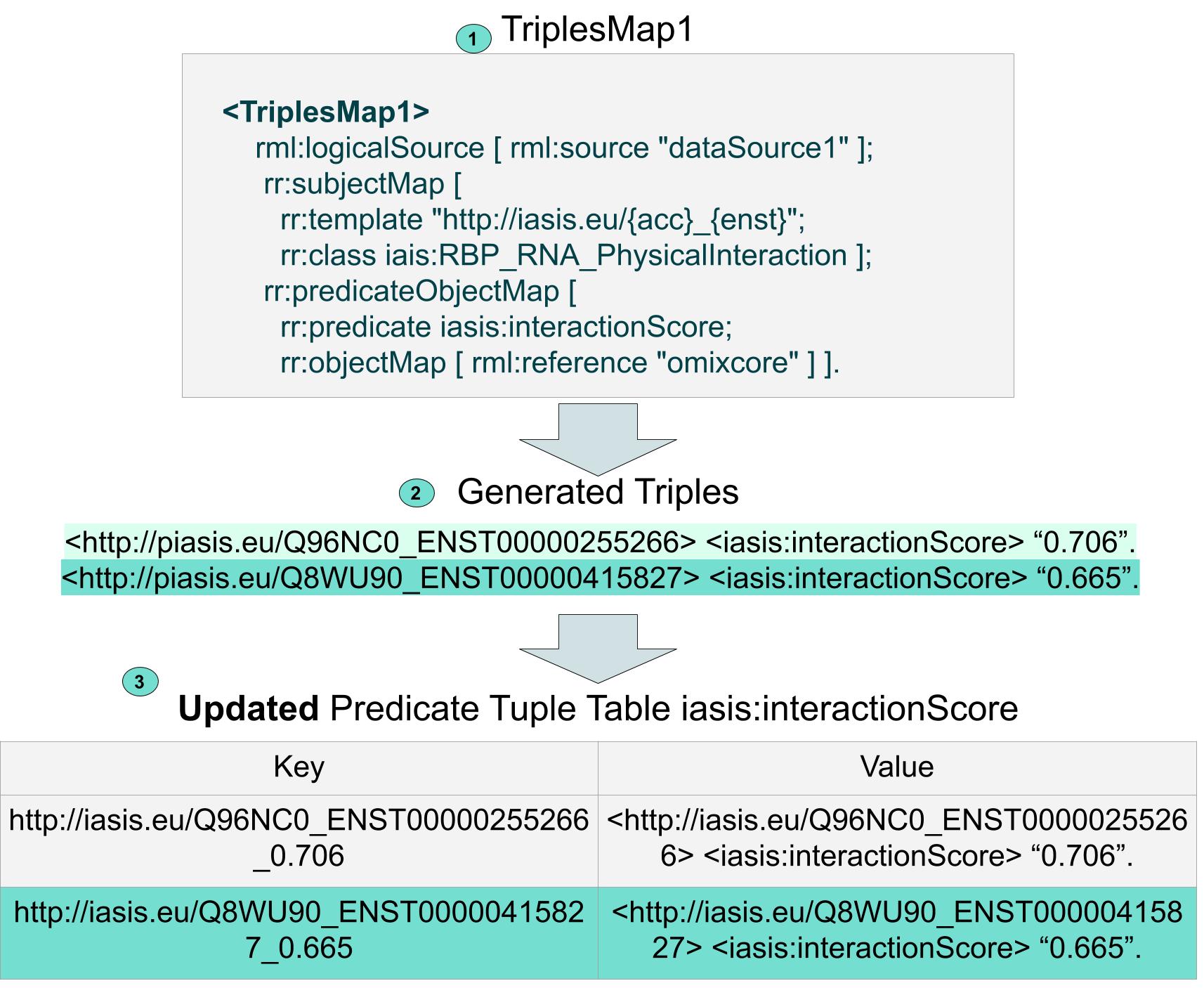}
    \label{fig:om}}
  \subfloat[Object Reference Map]{
\includegraphics[width=0.65\columnwidth,height=6.9cm]{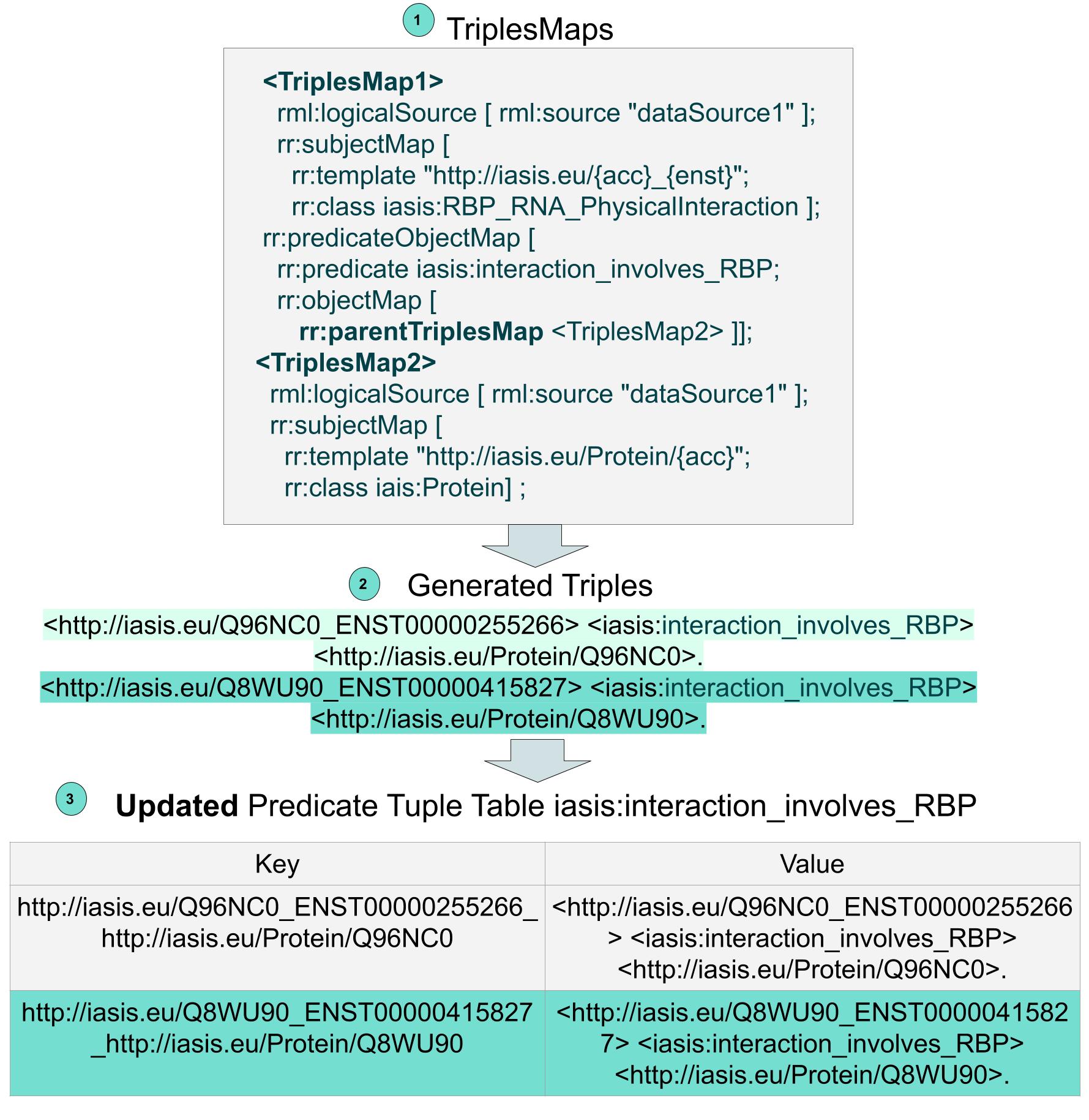}
    \label{fig:orm}}
  \subfloat[Object Join Map]{
\includegraphics[width=0.75\columnwidth]{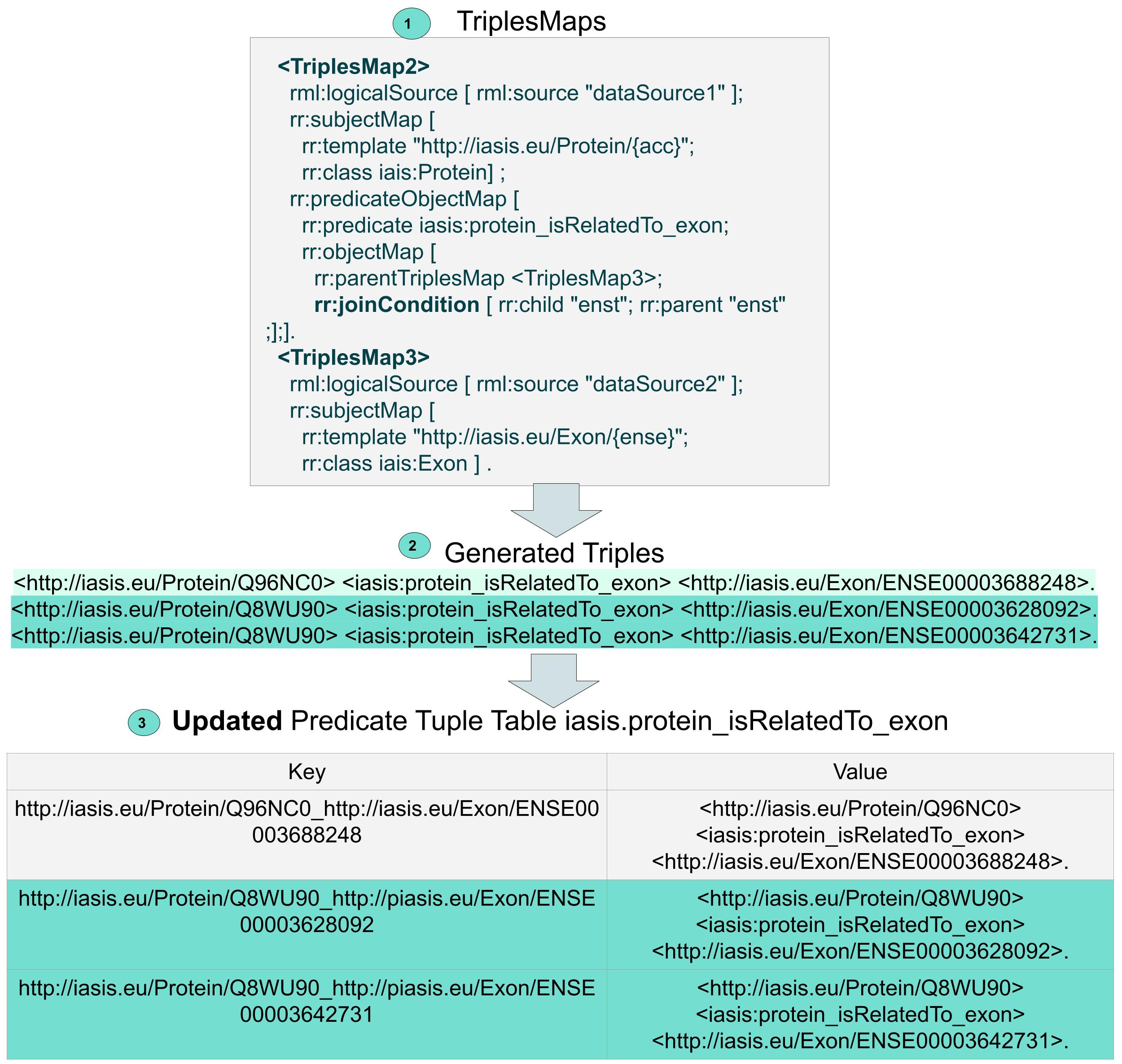}
    \label{fig:ojm}}
    \caption{SDM-RDFizer implements three physical RML operators that rely on PTTs to avoid the generation of duplicates. Object Join Maps resort to PJTTs to provide a direct access to the inner tables (i.e., the parent triples maps) of a join between two triples maps; also, PJTTs avoid the traversal of a parent triples map in case it is referenced by more than one triples map.}
    \label{fig:DTR}
\end{figure*}

\subsection{RML Operators and Algorithms}
\label{operators}
SDM-RDFizer implements three different operators for the creation of knowledge graphs. Depending on the type of the triples map, the SDM-RDFizer executes the respective operator. If the triples map has a join condition, then an \textbf{Object Join Map} operator is used. If the triples map has a reference to another triples map but does not have a join condition, then the \textbf{Object Reference Map} operator is used. Finally, if the triples map does not have a join condition or a reference to another triples map, then the \textbf{Simple Object Map} is used. We now describe the operators in more detail.

\noindent\textbf{Simple Object Map (SOM)}
 It is the most basic operator that SDM-RDFizer can execute and enables the generation of an RDF triple by performing a simple predicate object map statement. As illustrated in Figure\autoref{fig:om}, given a triples map and its respective data source, SDM-RDFizer generates RDF triples following what is established on the map. Each generated RDF triple is checked against the corresponding predicate tuple table (\textbf{PTT}). If the generated RDF triple already exists in PTT, then it is discarded. In the opposite case, the RDF triple is added both to PTT and the knowledge graph. This operation is depicted in Figure\autoref{fig:om}, where two RDF triples are generated. "<http://iasis.eu/Q8WU90\_ENST00000415827> <iasis:interactionScore> “0.665”." is not in PTT, then, it is added both to the table and the knowledge graph.  

\noindent\textbf{Object Reference Map (ORM)}
It seeks to expand what is established in Simple Object Map, by using the subject of a triples map as the object of another triples map. The primary condition for this operator to work is that both triples maps have the same data source. Afterward, the same process as in Simple Object Map is applied on the generated RDF triples, i.e., the triples are checked against PTT to determine if the triples are required for the knowledge graph creation. An example of this operation is in Figure\autoref{fig:orm}. There are two triples maps in the figure, where the \textit{<TripleMap2>} acts as the parent triples map. Two RDF triples are generated, but only the new one is included in the PTT. 

\noindent\textbf{Object Join Map (OJM)}
It seeks to expand what is established in Object Reference Map, but the main difference is that triples maps have different data sources. Using the corresponding PJTT, SDM-RDFizer implements an index join where the outer table of the join corresponds to the values in the child map and the inner table to the PJTT. Thus, to validate the satisfaction of a join condition $B$, the value of $B$ is checked in PJTT. If an entry $e$ exists with that hash key, all the subjects in $e$ are used to generate the resulting RDF triples. Finally, similar to the last two operations, the generated RDF triples are checked against the corresponding PTT to validate duplication and decide if they are going to be included in the knowledge graph. A way to better understand this operation is to view Figure\autoref{fig:ojm}. In the figure, the join condition is the column "enst" in both data sources. A PJTT table is created from the data associated with the join condition, as shown in Figure\autoref{fig:pjtt}. Three RDF triples are generated, and only two are not duplicates (i.e., they are not in the PTT). This operation is similar to Object Reference Map, since the object of the triples is the subject of the parent triples map. 
\subsection{Properties} 
We present the main properties of the RML operators implemented by SDM-RDFizer. Per operator \texttt{o}, we seek to compare the number of operations done by SDM-RDFizer versus the ones done by a na\"ive implementation of \texttt{o}; we named these expressions $\phi_{\texttt{o}}(.)$ and $\widehat{\phi}_{\texttt{o}}(.)$, respectively. Without lost of generality, we just focus on main-memory operations per operator, i.e., comparisons and insertions in main-memory data structures. Consider a predicate $p$, a multiset $N_p$, and set $S_p$; $N_p$ includes all the RDF triples of $p$ while $S_p$ is the corresponding set of $N_p$. Consider $|N_p|$ and $|S_p|$ as the cardinality of $N_p$ and $S_p$, respectively. In presence of a high-duplicate rate of RDF triples of $p$, $|S_p|$ is much smaller than $|N_p|$ (i.e., $|S_p| \ll|N_p|$).
\begin{itemize}
    \item \textbf{Simple Object Map (SOM):}
    Let $M$ be an RML triples map with an object triples map that defines $p$, $\phi_{\texttt{o}}(M)$ and $\widehat{\phi}_{\texttt{o}}(M)$ are defined as follows.  
  The na\"ive implementation of the simple object map operator $o$ in $M$ generates all the duplicates and then, it needs to execute a duplicate elimination process to add the RDF triples to the knowledge graph. Suppose a merge sort algorithm is conducted to eliminate duplicates \cite{BittonD83}\footnote{$\Theta(.)$ corresponds to the asymptotic notation}, then the following number of operations are required: 
       \[\widehat{\phi}_{\texttt{o}}(M)=
        |N_{p}| + |S_{p}| + \Theta(N_{p}log(N_{p}))\]
       
Contrary, the SDM-RDFizer algorithm of a simple object map resorts to a PTT of $p$ and never generates duplicates. As a result, the number of operations is defined as follows:
       \[ \phi_{\texttt{o}}(M)=|N_{p}| + 2|S_{p}|\]
       
    \item \textbf{Object Reference Map (ORM):} 
    This operator requires defining $p$, a reference of $M$ to a parent triple map $M_i$ expressed over the same data source $s$ of $M$. That is, the operator corresponds to a self-join over $s$ with a natural join condition on the attribute(s) that corresponds to the subject of $M$. As in a natural join, the join condition is not required. Assume $\Theta(1)$ is the cost of accessing the value of the subject of $M_i$ when $M$ is executed, then the number of operations is the same as executing a simple object map, i.e., 
    \[\widehat{\phi}_{\texttt{o}}(M)=
        |N_{p}| + |S_{p}| + \Theta(N_{p}log(N_{p}))\]
    \[ \phi_{\texttt{o}}(M)=|N_{p}| + 2|S_{p}|\]    
    \item \textbf{Object Join Map (OJM):}
    An Object Join Map executes a join between the data source of a child triple map $M$ and the data source of a parent triple map $M_i$ on a join condition $B$. In this case, $|N_p|$ represents the number of RDF triples resulting from evaluating the join and $|S_p|$ the number of duplicate-free RDF triples in $N_p$. Further, assume $|N_{\textit{parent}}|$ and $|N_{\textit{child}}|$ are the number of rows in the parent and child maps, respectively, to check to validate the join condition. If the na\"ive approach follows a nested loop join \cite{SteinbrunnMK97}, then 
    \[\widehat{\phi}_{\texttt{o}}(M)= |N_{\textit{parent}}| \times |N_{\textit{child}}| + \]\vspace{-16pt}
        \[|N_{p}| + |S_{p}| + \Theta(N_{p}log(N_{p}))\]
 Contrary, SDM-RDFizer relies on the PJTT $M_i \_B$ (of size $N_{\textit{parent}}$\footnote{We assume that a PJTT creation costs $N_{\textit{parent}}$ main-memory operations.}) and the PTT of $p$ to implement an index join that produces duplicate-free RDF triples. Thus, both physical data structures enable an efficient implementation of OJM. As a result, the number of operations is as follows:
 \[\widehat{\phi}_{\texttt{o}}(M)= 2|N_{\textit{parent}}| + |N_{\textit{child}}| +
        |N_{p}| + 2|S_{p}|\]
\end{itemize}
\begin{figure*}[t!]
 \centering
    \subfloat[10k records]{
        \includegraphics[width=0.7\columnwidth]{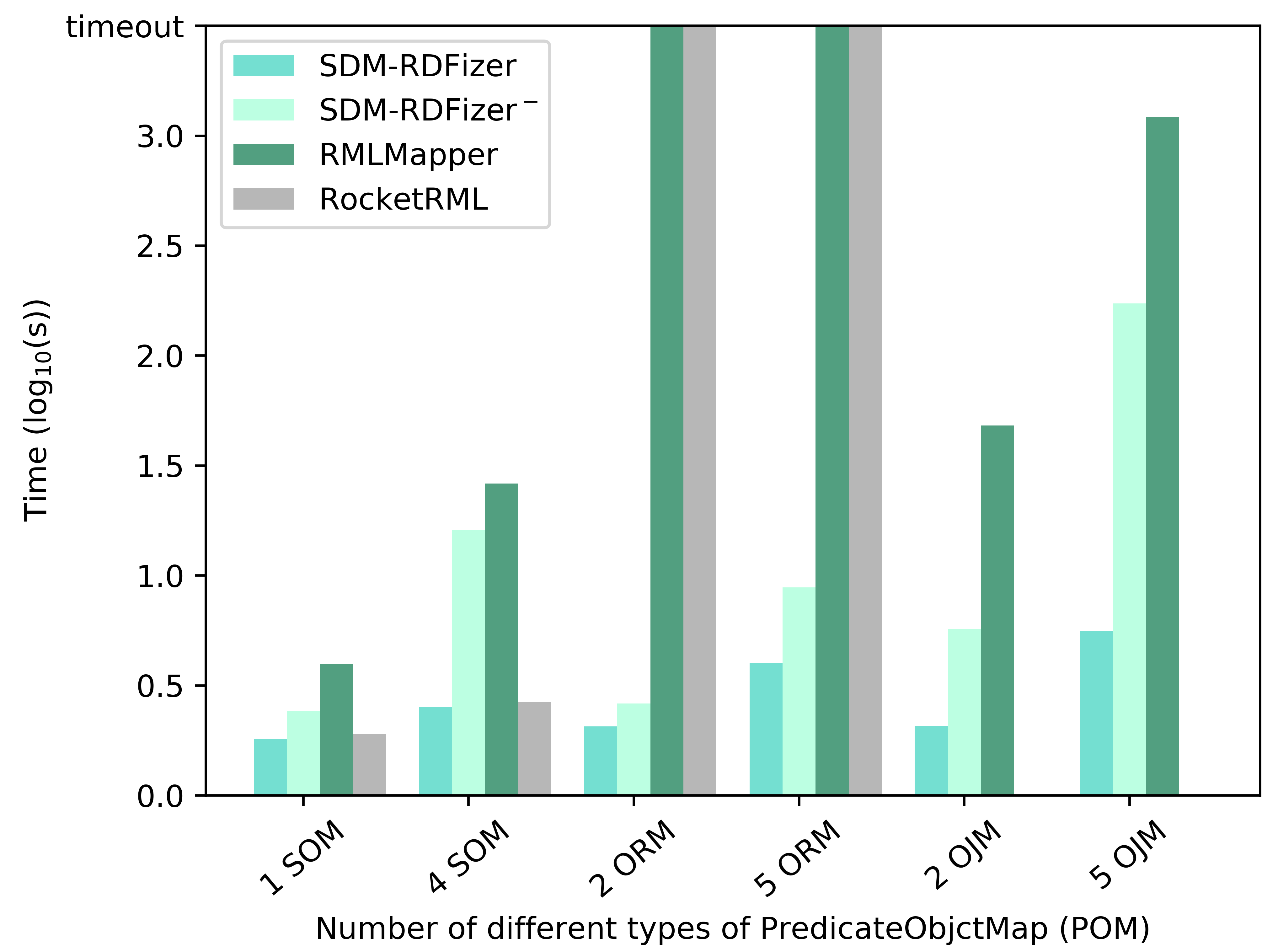}
            \label{fig:vera25_10K}}
    \subfloat[100k records]{
        \includegraphics[width=0.7\columnwidth]{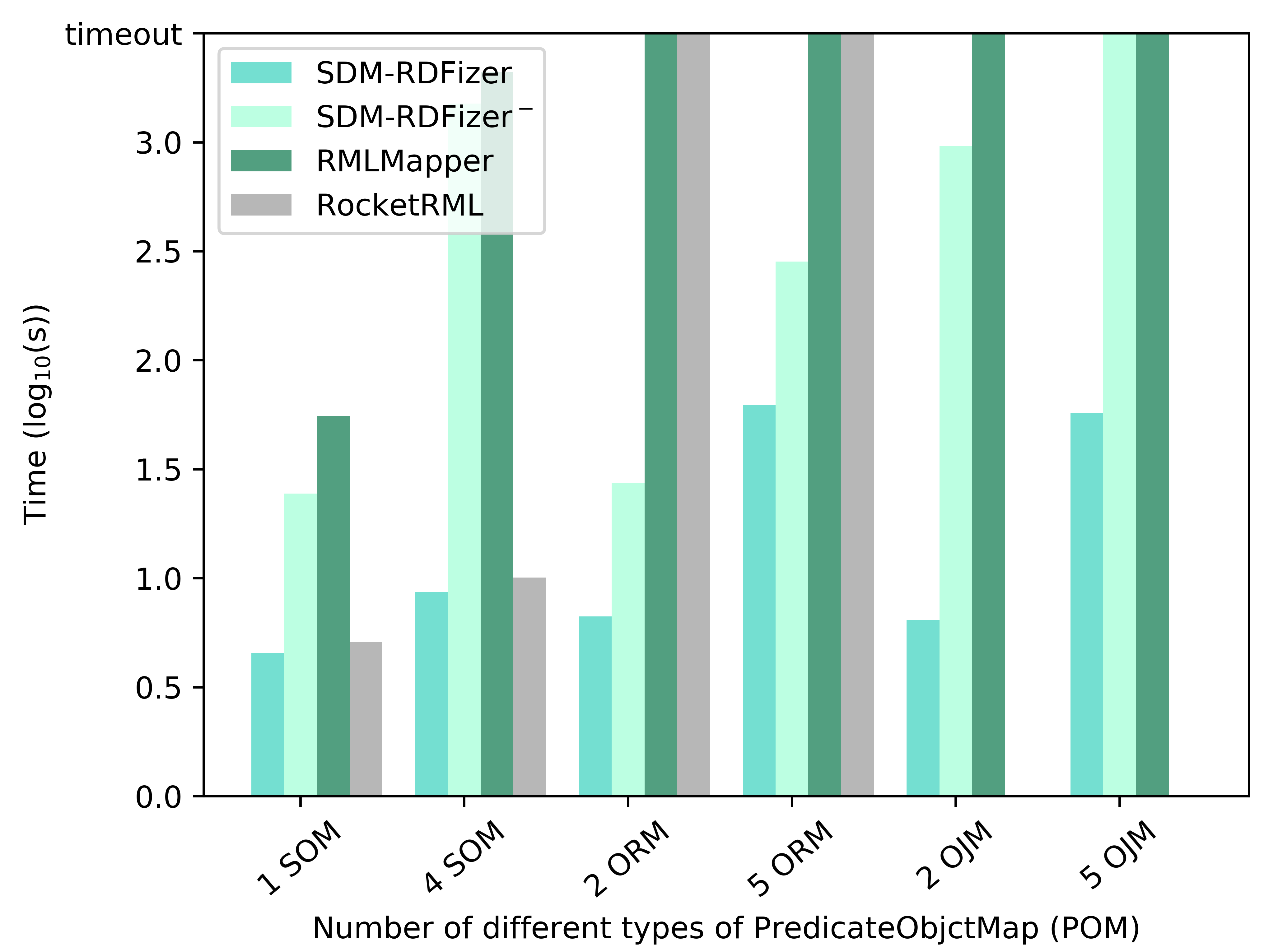}
            \label{fig:vera25_100K}}
 \subfloat[1M records]{
        \includegraphics[width=0.7\columnwidth]{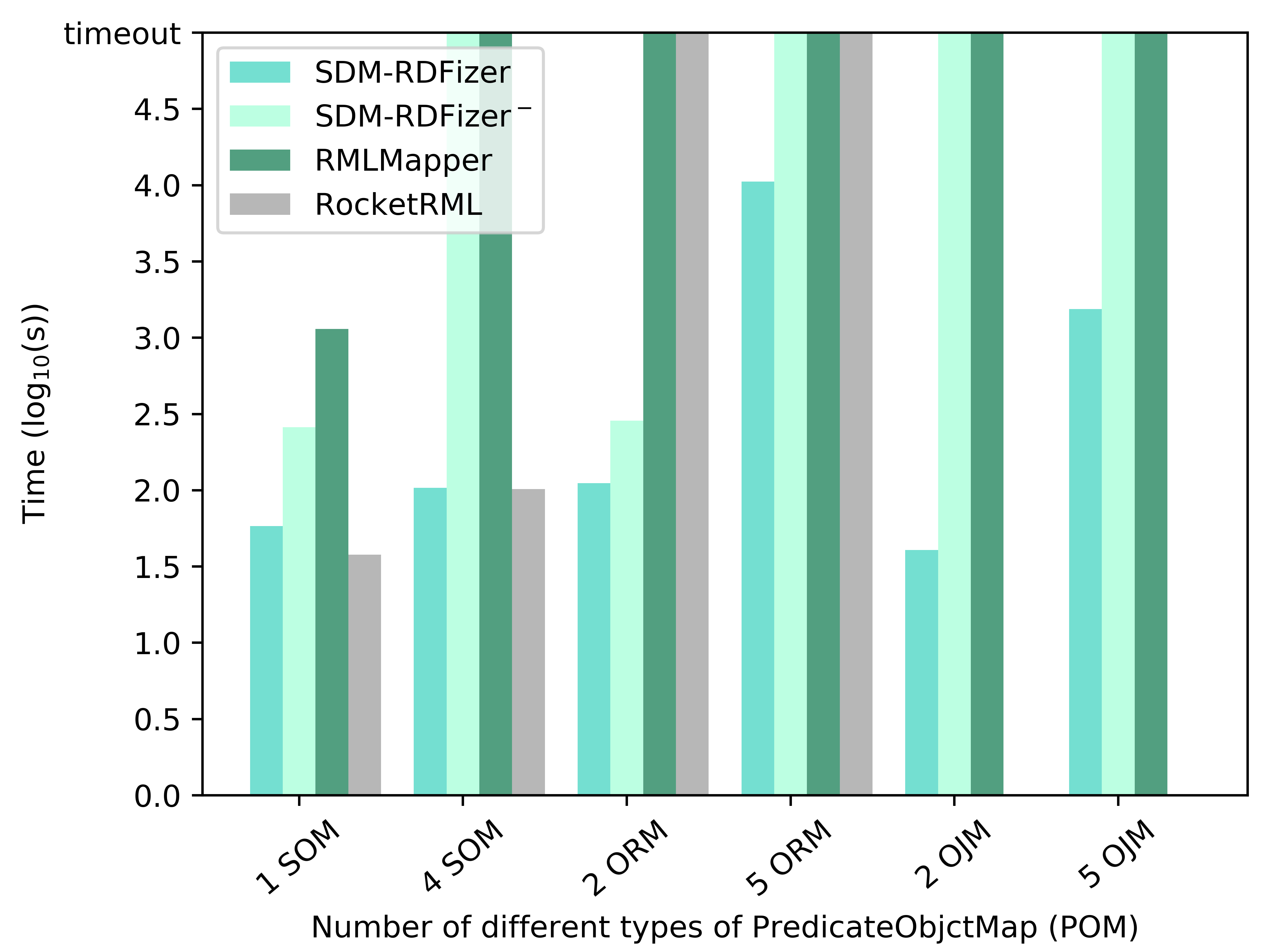}
            \label{fig:vera25_1M}}
\caption{{\bf Total execution time of experiments on datasets with 25\% duplicates.} SOM means simple object map, ORM object reference map and OJM object join map. RocketRML generates incorrect results running OJM mappings.}
    \label{fig:25percent}
\end{figure*}
\begin{figure*}[t!]
 \centering
    \subfloat[10k records]{
        \includegraphics[width=0.7\columnwidth]{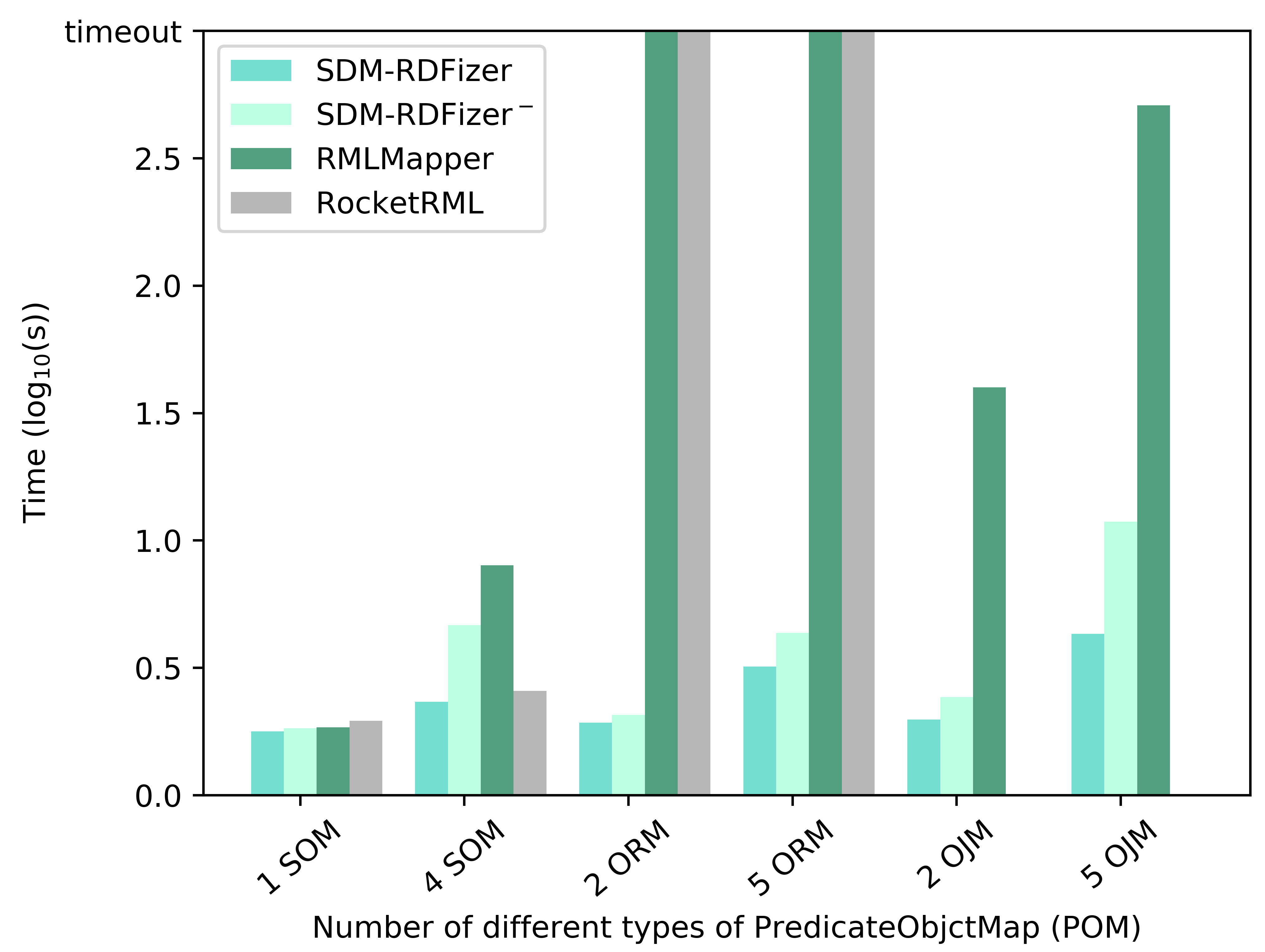}
            \label{fig:vera75_10K}}
    \subfloat[100k records]{
        \includegraphics[width=0.7\columnwidth]{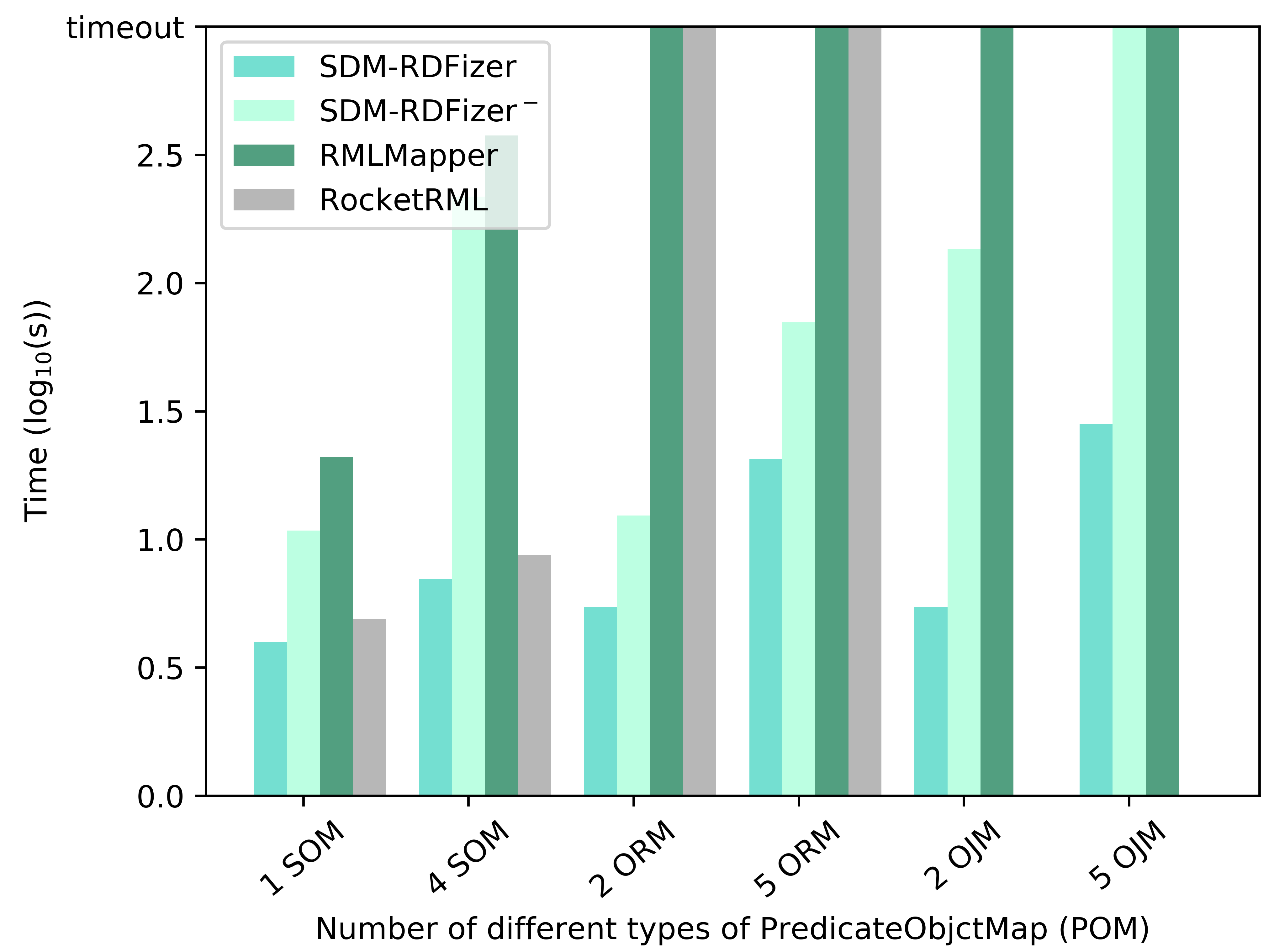}
            \label{fig:vera75_100K}}
    \subfloat[1M records]{
        \includegraphics[width=0.7\columnwidth]{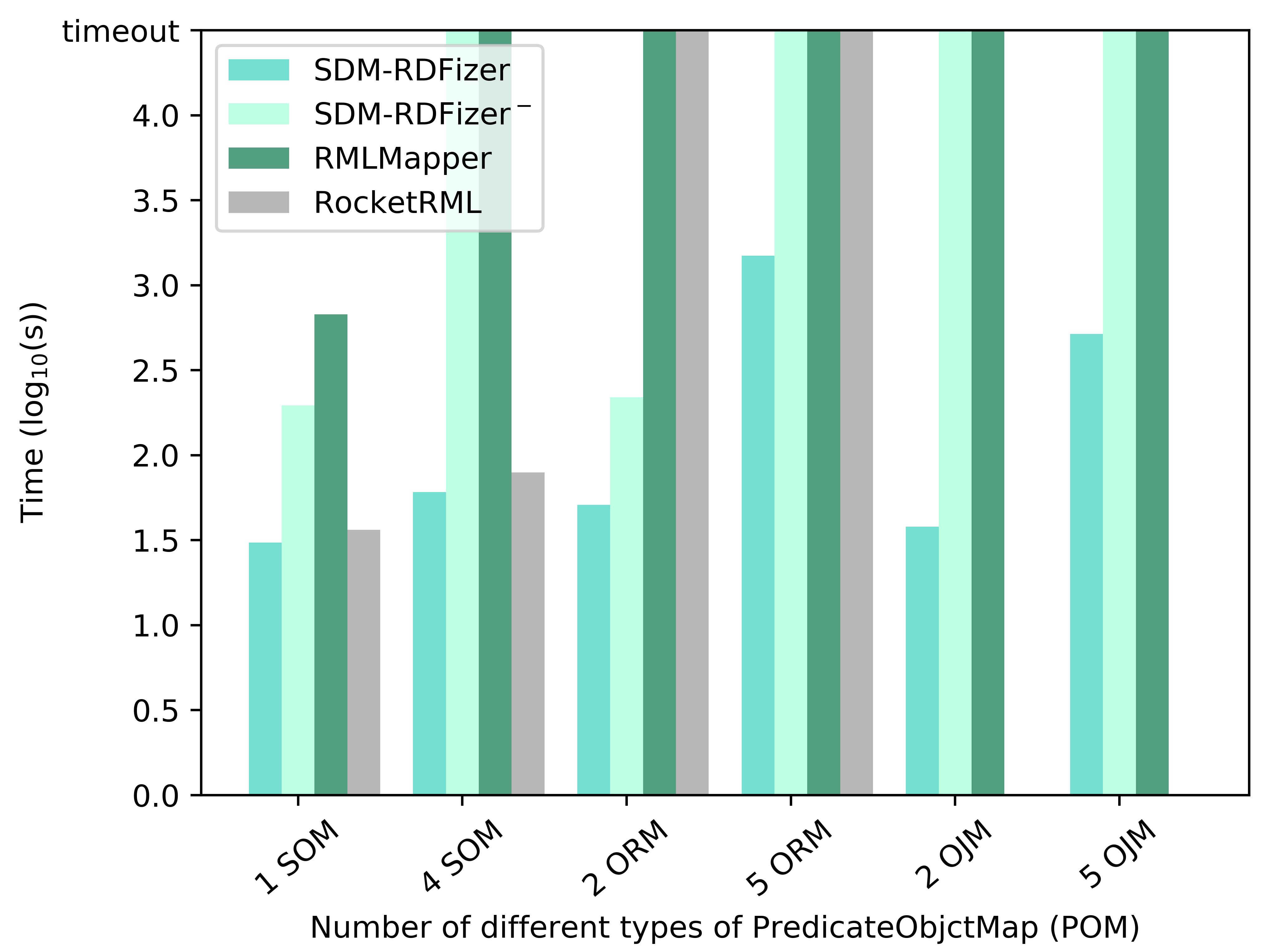}
            \label{fig:vera75_1M}}
\caption{{\bf Total execution time of experiments on datasets with 75\% duplicates.}  SOM means simple object map, ORM object reference map and OJM object join map. RocketRML generates incorrect results running OJM mappings.}
    \label{fig:75percent}
\end{figure*}
\section{SDM-RDFizer as a Resource}
\label{sec:resource}
\textbf{Novelty:}
SDM-RDFizer introduces a novel set of operators to execute mapping rules in a data integration system; they allow for efficient creation of knowledge graphs from heterogeneous data sources. Although the current version of SDM-RDFizer is customized for RML, the set of operators can be easily extended for other mapping rule languages and data models to represent knowledge graphs. The experimental studies comparing the performance of SDM-RDFizer illustrate the novelty of the proposed work with state of the art. We hope that these results encourage the community to advance existing approaches to scale up to the avalanche of available data that is expected in the next years.
\\
\textbf{Availability:}
SDM-RDFizer is released publicly by the Scientific Data Management (SDM) group at TIB, Hannover\furl{https://www.tib.eu/en/research-development/scientific-data-management/}. 
TIB is one of the largest libraries for science and technology in the world. Following its policy of engaging open access to scientific artifacts, it will keep available SDM-RDFizer as a tool for supporting the creation of knowledge graphs.  
The SDM-RDFizer is open source, written in Python 3, and available under the Apache License 2.0 license in an open Github repository\furl{https://github.com/SDM-TIB/SDM-RDFizer}; it is regularly updated with new features. 
Additionally, following open science good practices, we register the tool at the Zenodo platform, which takes the Github repository and gives a general DOI\furl{https://doi.org/10.5281/zenodo.3872103} to the engine and also a DOI for each release of the code\footnote{SDM-RDFizer v3.2: \url{https://doi.org/10.5281/zenodo.3872104}}. Thus, users and practitioners can use and cite a specific version of the engine, ensuring reproducibility and traceability of any experimental evaluation.
\\
\textbf{Utility:}
A docker image of SDM-RDFizer is available at DockerHub\furl{https://hub.docker.com/repository/docker/sdmtib/sdmrdfizer} and the Github repository of the resource, provides a detailed explanation of how to create and run the Docker container. Additionally, we provide a video demonstrating the steps required to execute SDM-RDFizer\furl{https://www.youtube.com/watch?v=DpH_57M1uOE}. The use case presented in the motivating example is utilized to facilitate understanding. Lastly, the activity of commits in the Github repository illustrates the attention paid to the creation of new versions and the resolution of the issues identified by the users of the tool.   
\\
\textbf{Predicted Impact:}
 The number of visits of knowledge graphs like DBpedia and Wikidata, and the current developments in scientific (e.g., \cite{AuerKPKSV18}) and industrial areas (e.g., \cite{NoyGJNPT19}) evidence the need of providing efficient tools for knowledge graph management at scale. The experimental evaluations of SDM-RDFizer illustrate the benefits of grounding solutions for the problem of knowledge graph creation in the well-established areas of data integration systems and query processing. Thus, we ambition that they will be the starting point of future developments. They include optimization techniques for enabling distributed mapping rule executions, as well as data integration systems capable of explaining the whole process of knowledge graph creation. 
 \\
\textbf{Adoption and Reusability:}
Several projects from different domains already use SDM-RDFizer.
iASiS\furl{http://project-iasis.eu/} and BigMedilytics - lung cancer pilot\furl{https://www.bigmedilytics.eu/} are exemplary of EU H2020 projects.
The iASiS RDF knowledge graph comprises more than 1.2B RDF triples collected from more than 40 heterogeneous sources using more than 1,300 RML triple maps. More than 800 RML triple maps are used to create, from 25 data sources, a lung cancer knowledge graph with 500M RDF triples. SDM-RDFizer has also created the \textit{Knowledge4COVID-19} knowledge graph during the EUvsVirus Hackathon\furl{https://devpost.com/software/covid-19-kg}; it comprises 28M RDF triples describing 63527 COVID-19 articles and related COVID-19 concepts (e.g., drug-drug interactions and molecular dysfunctions). SDM-RDFizer is also used in EU H2020, EIT-Digital, and Spanish national projects where the Ontology Engineering Group (Technical University of Madrid) participates. These projects, mainly focused on the transportation and smart cities domain, include: SPRINT\furl{http://sprint-transport.eu/}, SNAP\furl{https://snap-project.eu/} and Open Cities\furl{https://ciudades-abiertas.es/}. Similar to the \textit{Knowledge4COVID-19} knowledge graph, SDM-RDFizer also created the Knowledge Graph of the Drugs4Covid project\furl{https://drugs4covid.oeg.fi.upm.es/} where NLP annotations and metadata from more than 60,000 COVID-19 articles are integrated into almost 44M RDF triples.

\section{Empirical Evaluation}
\label{sec:eval}
We compare SDM-RDFizer with a baseline and existing RML interpreters. We aim to answer the following research questions:\textbf{Q1)} What is the impact of data duplication rate in the execution time of a knowledge graph creation approach? \textbf{Q2)} What is the impact of input data size in the total execution time of a knowledge graph creation process? \textbf{Q3)} What is the effect of the triples map types in the \verb|PredicateObjectMap| of an RML mapping affect the existing engines? All the resources used in this evaluation are publicly available\furl{https://github.com/SDM-TIB/SDM-RDFizer-Experiments}. The experimental configuration is as follows:

\noindent\textbf{Datasets and Mappings.} To the best of our knowledge, there is no testbeds to evaluate the performance of a KG creation approach from heterogeneous data sources. Consequently, following the real-world scenario that initially motivated this research, we create our testbed from the biomedical domain. From the coding point mutation dataset in COSMIC\footnote{\url{https://cancer.sanger.ac.uk/cosmic} GRCh37, version90, released August 2019}, we randomly select records to create six datasets with different sizes, i.e., 10K, 100K, and 1M number of rows. Accordingly, each two datasets with the same volume size differ from each other in the number of duplicated values; including 25\% or 75\% of duplicates with each duplicated value to be repeated 20 times. In total, three mapping files are created with different types of \verb|PredicateObjectMap|: Simple Object Map rules with reference to columns (SOM), Object Reference Map rules (ORM), and Object Join Map rules (OJM). Each type of rules also varies from 1 to 4 number of \verb|PredicateObjectMap|.

\noindent\textbf{Engines.} The SDM-RDFizer v3.2 is tested in two different configurations: optimized version including the proposed operators (SDM-RDFizer) and the baseline with the na\"ive operators (SDM-RDFizer$^-$). Additionally, we also run the experiments over two well-known RML-compliant engines: RMLMapper v4.7\furl{https://github.com/RMLio/rmlmapper-java} and RocketRML v1.7.0\furl{https://github.com/semantifyit/RocketRML/}. A docker image is available per tested engine to facilitate reproducibility of the study.

\noindent\textbf{Metrics.} \textit{Execution time:} Elapsed time spent by an engine to complete the creation of a knowledge graph; it is measured as the absolute wall-clock system time as reported by the \verb|time| command of the Linux operating system. \textit{Number of RDF triples} in the knowledge graph. Each experiment was executed five times and average is reported. The time out is set to 5 hours. The experiments were run in an Intel(R) Xeon(R) equipped with a CPU E5-2603 v3 @ 1.60GHz 20 cores, 64GB memory and with the O.S. Ubuntu 16.04LTS.

\subsection*{Discussion}
In this section, we describe the outcomes of our experiments evaluating the performance of the selected engines (i.e., SDM-RDFizer, RMLMapper, and RocketRML) in different testbeds.
\autoref{fig:25percent} and \autoref{fig:75percent} report on execution time for creating a knowledge graph from datasets with 25\% and 75\% of duplicates, respectively. It should be noted that since RocketRML does not support N-M join relations and generates incorrect outputs; subsequently, we only provide the results of SOM and ORM mappings for this engine. In all the experiments, we have verified that the generated outputs are the same for all the approaches. \\
\noindent The results reveal the benefits of applying the proposed operators during the creation of a knowledge graph. As illustrated in \autoref{fig:25percent} and \autoref{fig:75percent}, independent of the size of the input datasets and the percentage of existing duplicates, RMLMapper and RocketRML fail to generate RDF triples from mappings including 2-ORM and 5-ORM; they time out in five hours. Moreover, the execution time of RMLMapper and RocketRML increases as the size of data and number triples maps increase. Nonetheless, as observed, SDM-RDFizer completes the RDF triples generation in all testbeds within a reasonable period. Additionally, the performance of SDM-RDFizer$^-$ provides evidence of the quality of the SDM-RDFizer operators and their ability to speed up a knowledge graph creation process.       
\section{Related Work}
\label{sec:rw}
Solutions provided to the problem of knowledge graph creation from structured and semi-structured data are gaining momentum in practitioners and users~\cite{corcho2020towards}. Since 2012, with the appearance of the W3C recommendation of the R2RML~\cite{das2012r2rml} mapping language for transforming relational databases to knowledge graphs, diverse proposals, and optimizations techniques have tackled this problem. For example, \cite{gawriljuk2016scalable} presents a framework for incremental knowledge graph creation. Additionally, diverse optimizations techniques are proposed for virtual knowledge graph generation\cite{calvanese2017ontop,PriyatnaCS14}, i.e., SPARQL-to-SQL query translation techniques~\cite{chebotko2009semantics}. Albeit efficient, these proposals are only focused on relational database data sources, and most of the optimizations are done over a SQL translated query that cannot be applied when datasets are heterogeneous.

With the RML~\cite{DimouSCVMW14} specification proposed as an extension of R2RML for describing mapping rules over diverse data formats, multiple tools have been developed for the creation of knowledge graphs. For example, CARML\furl{https://github.com/carml/carml} executes RML rules and includes additional features like MultiTermMap (to deal with arrays) and XML namespace (to improve XPath expressions). RocketML~\cite{csimcsek2019rocketrml} is a RML engine implemented using the NodeJS framework and RMLMapper is also a commonly used RML engine. Unfortunately, none of these engines implements efficient operators to execute mapping rules. As a result, they cannot scale up to a large volume of data, high-duplicate rate, and heterogeneity. Our ambition is to provide a resource for generating knowledge graphs at scale.
\section{Conclusions}
\label{sec:conclusion}
The observation that both industrial and scientific applications demand efficient solutions for knowledge graph creation motivated the need to make SDM-RDFizer available as a resource. SDM-RDFizer implements novel physical operators and data structures that speed up the generation of duplicate-free RDF triples even in the presence of data sources with a high-duplication rate. Empirical results indicate that SDM-RDFizer outperforms state of the art by up to three orders of magnitude. Thus, SDM-RDFizer broadens the portfolio of technologies for knowledge graph management and provides the basis for developing real-world knowledge graph applications. In the future, we will devise optimization techniques to plan the execution of the mapping rules and extend SDM-RDFizer to other mapping languages.


\bibliography{sdm-preprint}

\end{document}